# The Information Theory of Emotions of Musical Chords

**Vadim R. Madgazin**     mailto:vrm@vmgames.com

*The paper offers a solution to the centuries-old puzzle - why the major chords are perceived as happy and the minor chords as sad - based on the information theory of emotions.*
*A theory and a formula of musical emotions were created. They define the sign and the amplitude of the utilitarian emotional coloration of separate major and minor chords through relative pitches of constituent sounds [15].*
*Keywords: chord, major, minor, the formula of musical emotions, the information theory of emotions.*

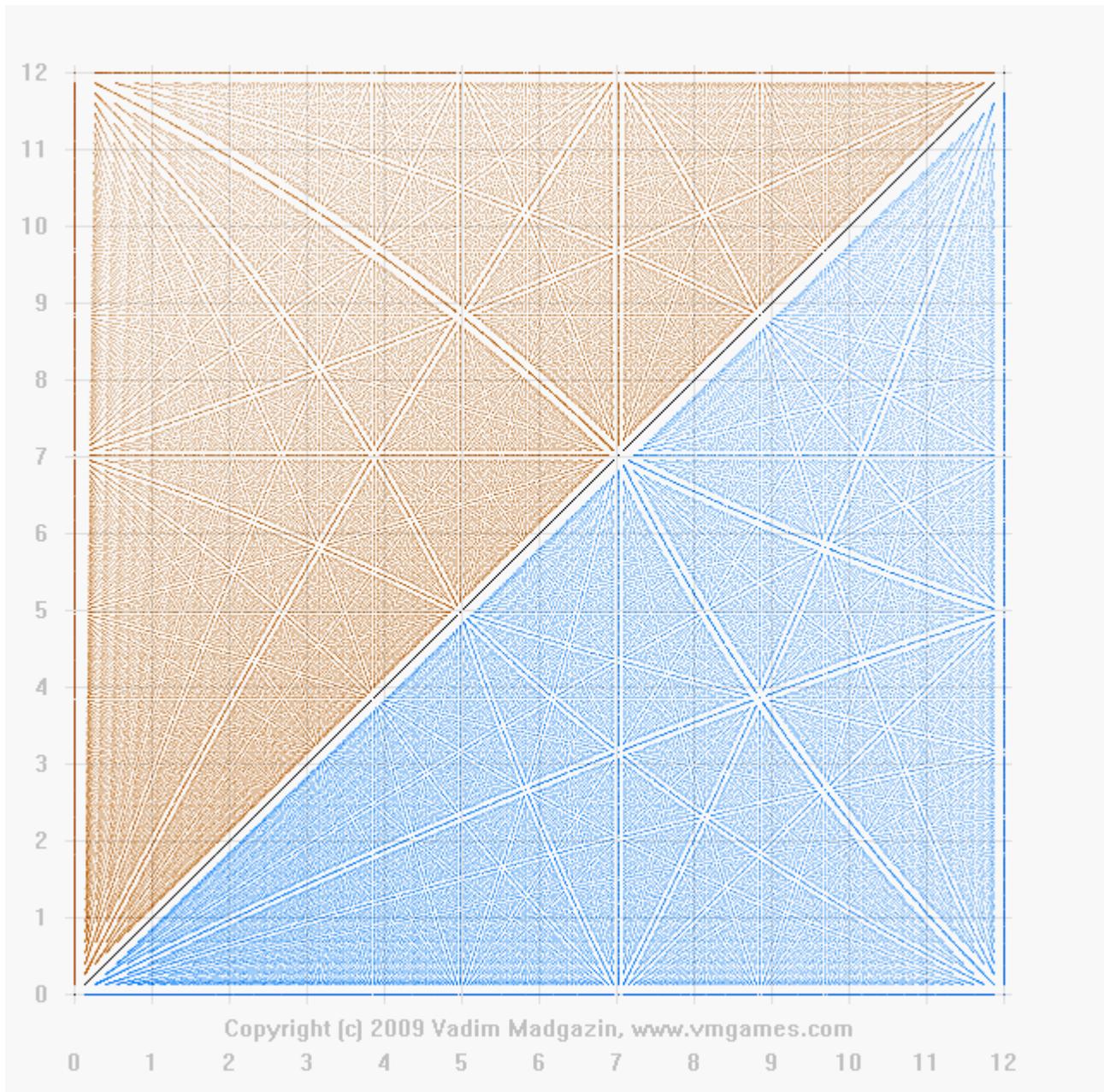

Figure 1. Major and Minor triads space of octave.
Intensity of point = const/log(n1*n2*n3) where n1, n2, n3 = 1, 2, 3, ...130

# 1. CHORD PROPORTIONS

We shall study the psychological effects from listening to several musical sounds taken simultaneously outside of any musical context. Our attention will be concentrated on the ratio of pitches of the principal tones of sounds in a separately taken chord. We call this ratio of pitches a proportion.

Depending on a proportion, a dyad (two sounds) can produce either a pleasant (consonance) or an unpleasant (dissonance) aesthetic impression. It is considered, that consonance proportions consist entirely of numbers 1,2,3,4,5,6, and 8.

A triad (a chord of three sounds) possesses a new quality. It is the ability to cause various emotions in the subject. These "musical" emotions are subdivided into aesthetic and utilitarian ("sadness" and "happiness") [11, 12].
The emotions caused by separate chords do not depend on common pitch modifications, loudness or the quality of their component sounds. Further, we shall study utilitarian emotions generated primarily by major and minor triads.

A major triad (proportion 4:5:6, is approximately realized by notes Do, Mi, and Sol, or C-E-G) corresponds to positive emotions of "happiness" while a minor triad (proportion /4:/5:/6 with omitted unities in numerator, notes Do, Mi-flat, and Sol, or C-Eb-G) corresponds to negative emotions of "sadness". Emotions of all other major and minor chords always have the same corresponding signs.

Mathematically, any triad consists of three dyads and one common triple proportion. Using triples of coprime numbers *A,B,C* and *D,E,F,* a triple proportion can be written as a "direct" proportion (*A:B:C) or as an "inverse" proportion (/D:/E:/F).*

These proportions of triads can be divided into three groups:
1. Direct proportion is smaller          $A*B*C < D*E*F$
2. Inverse proportion is smaller         $A*B*C > D*E*F$
3. Both proportions are identical        $A*B*C = D*E*F.$

Absent in dyads, the new (emotional) quality of triads can be produced only by these common proportions that fall into one the three groups described above.

Consonance of triads usually does not exceed consonance of their constituent dyads. In musical practice, there are four main types of triads: major and minor (consonance), augmented and diminished (dissonance). Consonance triads primarily belong to the first two groups of proportions.

The direct and inverse proportions of augmented and diminished triads are identical, and therefore they belong to the third group:
/25:/20:/16 = 16:20:25
/36:/30:/25 = 25:30:36.

All major triads belong to the first group while all minor triads belong to the second group.
If a triad *A:B:C* is major, then inverse triad */C:/B:/A* is minor (though all numbers in the triads are identical, relative pitches of their middle sounds will be different). Each two of such triads form a complementary pair of chords consisting of identical dyads.

Any direct proportion can always be presented as an inverse proportion, and vice versa:
4:5:6 = /15:/12:/10
/4:/5:/6 = 15:12:10.

Taking into consideration that the human ear is rather advanced [8, 9, 14], it is possible to assume that if the subject's higher nervous system is quite capable of perceiving a minor triad as a direct proportion (15:12:10), it is also capable of perceiving the same triad as an inverse proportion (/4:/5:/6) and indentifying this triad as belonging to the minor second group. Similar classification can be applied to any other consonance triad as well as to a number of relatively simple dissonance triads.

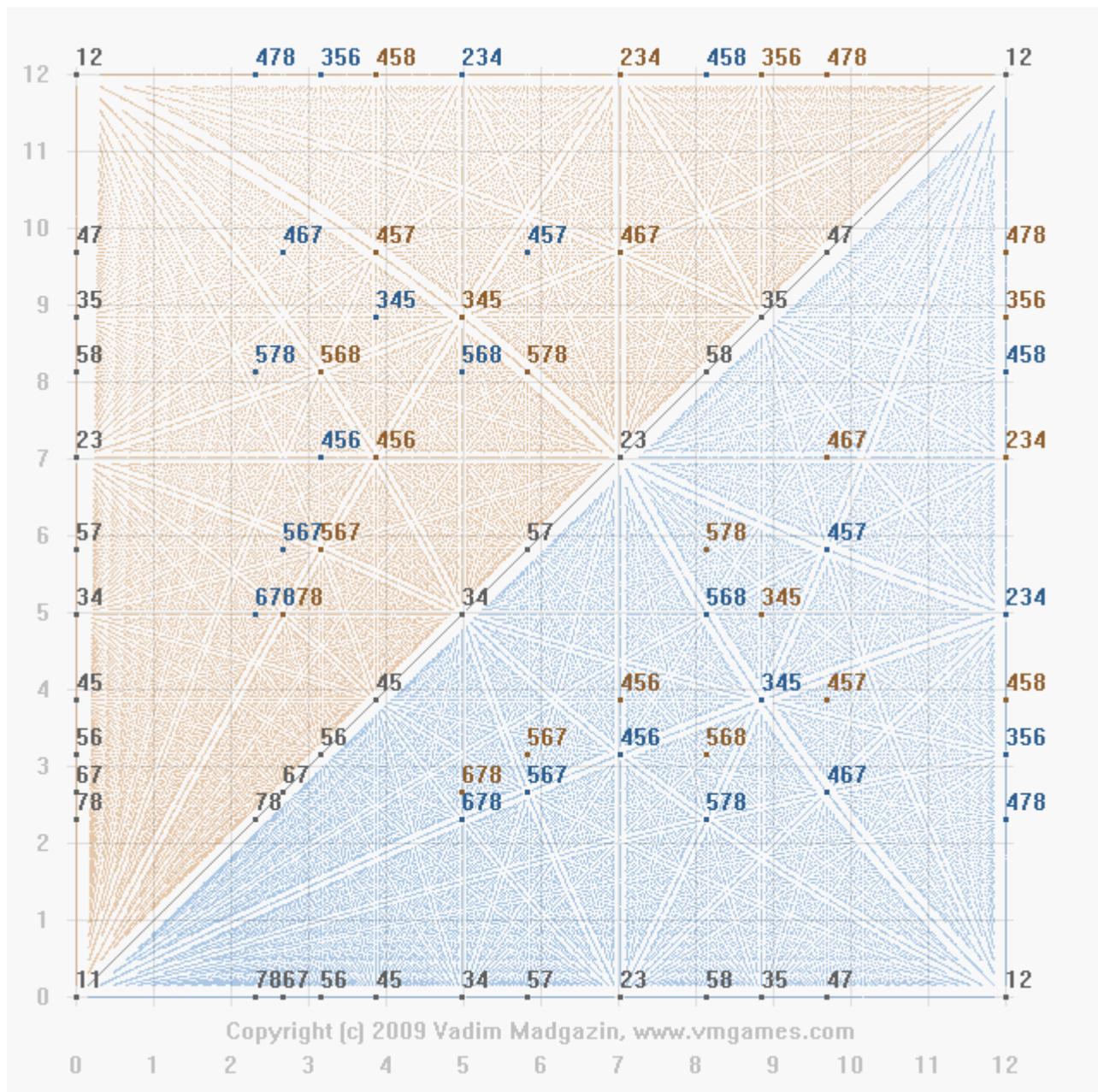

Figure 2. Major and Minor triads and Diads.

<u>We shall further call as the main proportion of a chord the one of the two proportions (direct or inverse) the multiplication of the numbers of which makes a lesser value. And we shall call the other proportion as the side proportion.</u>

A direct proportion of a major triad and an inverse proportion of a minor triad will always be the main proportion. The emotional "information" of a chord should be uniquely determined by the type of its main proportion. There is simply no other source of this information that must be present in a chord. There is not and cannot be any other source, since we consider stationary chords with a constant spectrum, in the extreme case consisting of only pure tones.

## 2. THEORIES OF MAJOR AND MINOR

As early as 1558, Gioseffo Zarlino knew the opposite meaning of major and minor chords [1]. However, over the past 450 years the theory of music has not advanced far from this knowledge.

Here is a phrase from a doctoral thesis on musicology written in 2008 expressing apparently the bottom line of our theoretical knowledge of major and minor [13, page 91]:
"Although many authors have written about the perception of major/minor chords and scales, it still remains a riddle why major chords are felt as happy and minor chords are felt as sad."

Nevertheless, we should not forget history at all. The idea to search for the "meaning" of a chord outside of the music' "own space" was proposed for the first time at least over a hundred years ago.

Here is a quote [3, 4]:
By the end of his active work, Hugo Riemann renounced the validation of major and consonance by means of overtones and took sides with Stumpf's psychological point of view considering overtones only as "an example and confirmation" but not as the evidence.
Carl Stumpf transferred the scientific validation of the theory of music from the area of physiology to the area of psychology. Stumpf refused to explain consonance as an acoustic phenomenon and described it as a psychological fact of "tonal fusion".

It took psychology a whole long century before it could reach a certain level of sophistication and start using appropriate math tools. As is known, the level of precision of any science depends on the amount of mathematics it uses.

Now we must approach the problem of major and minor "from the other end" and ask a question: what is an emotion?

## 3. FORMULAS OF EMOTIONS

As of today, there is only one theory of emotions the language of which is capable, I hope, of mathematically describing the structure of psychological phenomena of music perception.
It is the information theory of emotions [5, 6, 7, 10].

Dependence of an emotion on its objective parameters is called the formula of emotions. Let's assume that the subject experiences a need in something by the magnitude of *W*. If he is lucky to receive a useful resource *R* that satisfies this need, his emotion *E* will be positive, and in the case of loss of *R* the emotion will be negative [7]:

*E = F(W, R)*   (1)

For example, let's imagine a gambler.
If the gambler won a certain sum $R>0$, he experiences a positive emotion of happiness by the amount of $E$. If the gambler "won" the sum of $R<0$ (i.e. he lost), this causes a negative emotion of sadness by the amount of $E$.

Another approach considers emotions as a means of optimal control of behavior directing the subject to reaching the maximum of his "objective function" $P$ [10].
The augmentation of function $P$ is accompanied by positive emotions and diminution - by negative emotions.

Since $P$, in the elementary case, depends on variable $x$, the emotion $E$ is caused by the change of this variable over time [10]:

$E = dP/dt = (dP/dx)*(dx/dt)$   (2)

Let's note, the formulas (1, 2) are somewhat similar, if we take into account that parameter $R$ in (1) is actually a difference between the current and the previous value of some integral useful resource $S$ that is proportional to the objective function of the subject.
For example, in the case of our gambler, it is logical to select his total capital as the quality of $S$, then $R=S1-S0=dS=dP$.

Let's improve the formulas of emotions by writing them in relative values. In fact, the happiness of the gambler should be proportional to the relative amount of the prize, instead of the absolute. Taking into account the Weber-Fechner law, we can also assume that the value of experience ($E$) is proportional to a logarithm of some "generative stimulus".

Let's write the modified formula of emotions as follows:

$E = F(W)*log(S1/S0)$,   (3)

Here, $F(W)$ is separately taken dependence of emotion on the parameter of need $W$,
$S1$ – the current value of objective function $S$ (or the total useful resource), and
$S0$ – the value of $S$ in an instant preceding the situation that caused the emotion.

Let's divide both parts (3) by $F(W)$ and obtain "specific power of emotions per unit of need" $Pwe=E/F(W)$. Let's also introduce dimensionless quantity $L=S1/S0$ that should be logically called a relative objective function. Then:

$Pwe = E/F(W) = log(S1/S0) = log(L)$,   (4)

Brief title *Pwe* means "power of emotions". *Pwe* is proportional to the emotional energy consumption per unit of time, i.e. it is identical to the common expression "force of emotions". The measurement of the force of emotions in terms of power released by an organism for an emotional behavior of the subject was proposed in [5].

It is easy to notice that formula (4) produces the correct sign for emotions: it is positive for the increase of $S$ (when $S1>S0$ and $L>1$) and negative for the decrease of $S$ (when $S1<S0$ and $L<1$).

Let's apply the new formula of emotions (4) to perception of musical chords.

# 4. THE INFORMATION THEORY OF CHORDS

I assume that certain rather simple analogies (at the level of coincidence of valuations, such as more/less, better/worse) with the meaning of similar information from other sensory channels of perception (visual, etc.) allow the mind of the subject to classify major chords as signals carrying information about increase/profit/success accompanied by positive emotions, and minor chords as signals carrying information about decrease/loss/misfortune accompanied by negative emotions.

I.e., using the language of the formula of emotions (4), a major chord must contain information about an event-relative objective function $L>1$, and a minor chord - about $L<1$.

My basic hypothesis is as follows. While perceiving a separate musical chord, the value of a relative objective function $L$ that is directly related to the main proportion of pitches of its constituent sounds is generated in the mind of the subject. The idea of an increasing value of the objective function ($L>1$) accompanied by positive utilitarian emotions corresponds to a major chord, while the idea of a decreasing value of the objective function ($L<1$) accompanied by negative utilitarian emotions corresponds to a minor chord.
If the main and the side proportions of a chord are identical, then this chord is neither major nor minor and the utilitarian component of its emotions is absent.

According to this hypothesis, the triads from the third group differ from those of the first and the second groups (major and minor) by the character of the emotional response of the subject - they cause only aesthetic emotions that have no sign.

Let's write formula (4) for an arbitrary number ($M$) of voices of a chord. For this purpose, we shall define $L$ as geometric average of all the numbers of the main proportion of a chord. Hence, we get the final version of "the formula of musical emotions":

$Pwe = log(L) = (1/M)*log(n1*n2*n3* ... *nM)$,   (5)

Here, $ni$ is the $i$-th voice in the chord in the form of a corresponding integer number or an inverse fraction from the main proportion of pitches of a chord, and the logarithm has the base of 2.

It follows from formula (5) that $Pwe$ of a major triad and $Pwe$ of the complementary minor triad will coincide in amplitude and differ only in the sign.

The logical name for $Pwe$ (5) would then be the "emotional power of a chord", or simply the "power of a chord", positive for major and negative for minor (the analogy would be: inflow and outflow of emotional energy).

# 5. DISCUSSION OF RESULTS

So, by putting forward a number of rather simple suppositions, we have obtained new formulas (3, 4, and 5) that link generalized parameters of a situation (or specific parameters of chords) with the sign and the force of utilitarian emotions caused by them within the context of a situation.

How can this result be evaluated?
Here is a quote [5]:
"Probably there have been no attempts to objectively define the force of emotion. However, it is possible to assume that such a definition should be based on power representations. If an emotion causes some type of behavior, this behavior requires certain energy spending. The stronger the emotion is, the more intensive is the behavior and the more energy per time unit is required.
I.e., it is possible to try identifying the force of emotion with the quantity of power that the organism spends on respective behavior."

Let's try taking the most critical approach to our new results, since they cannot be compared with anything else so far.

First, though the power of emotions *Pwe* from formulas (4, 5) is proportional to the "subjective force" of emotions, their relation may not be linear. This relation is only a certain average dependence across the entire continuum of the subjects and can be subject to individual deviations. For example, it is possible that instead of geometric average in formula (5) it is necessary to apply some other function of the numbers that make up a chord proportion.

Second, when we consider a certain variant of the formula of musical emotions (5), it should be noted that, though *M* in this formula may formally be equal to 1 or 2, we can rightfully talk about the emergence of utilitarian emotions only if *M*>2. However, already if *M*=2 we observe the emergence of aesthetic emotions, and when *M*>3, additional factors may possibly appear that may in some way influence the result.
In the case of *M*=2, formula (5) very manifestly, up to factor 1/2, coincides with the harmonic distance function proposed in [16] for the analysis of musical intervals.

Third, apparently, the area of valid amplitude values of *Pwe* for the category of major and minor has the upper bound of 2.7...3.0. However, already at values of 2.4 and above, the area of saturation of utilitarian-emotional perception of chords begins, and approximately at these values we observe the lower bound of the range of possible dissonance "intrusion."
The latter is a common problem of "non-monotonicity" of a series of dissonance intervals and it is not directly connected with the emotional aspect of perception of chords. For the power of emotions, the existence of dynamic range bounds is a common property of any sensory system of humans. It may be easily explained by the absence of "real life" analogies corresponding to dramatic changes in the value of the objective function, such as 10 times and more for a single event.

Fourth, the chords with direct and inverse proportions consisting of identical (or almost identical) numbers do not possess (or almost do not possess) the utilitarian component of emotions, which corresponds to zero (or close to zero) *Pwe*. In this case, however, formula (5) will produce a different, incorrect result.
This is a manifestation of a rather simple phenomenon: since it is impossible to differentiate between direct and inverse proportions of a chord by their complexity, the subject cannot determine the direction of the change in the value of the objective function and the classification of this chord in the categories of utilitarian emotions is simply not performed.

The formal result of the formula (5) usage can be supplemented by an additional rule. Two values are determined for each chord: *Pwe* for the main proportion and *Pwe2* for the side proportion of a chord.
If *Pwe* and *Pwe2* are close in amplitude, utilitarian emotions will be proportional to their half-sum (i.e., almost zero):

*E = (Pwe+Pwe2)/2*,   (5a)

This rule apparently begins to work when module (*Pwe* + *Pwe2*) is smaller than 0.50.

However, no matter what imaginary or real defects these formulas (3, 4, 5, and 5a) may have, they still give us a useful material for numerical valuations of the force of emotions, which are rather simple to obtain at least in one specific area, in the area of emotional perception of separate major and minor chords.

Let's listen to major triads 2:3:5 and 3:4:5. My ear can definitely distinguish that *Pwe* of the first triad (1.64) is really somewhat smaller than *Pwe* of the second triad (1.97).
Something similar can be observed with respective pairs of minor triads.

The same may be tried with any other pair of consonance triads (from the 1st or the 2nd but not from the 3rd group). In my opinion, the result is similar: the power (force, brightness, depth) of emotions seems to be greater for the triad with a greater amplitude of *Pwe*.

This impression of proportionality of the force of emotions and *Pwe* is also preserved when listening to the chords composed of sounds with a rich harmonic spectrum.

## 6. CONCLUSIONS

The goal of my work was to establish a possible nature of emotional coloration of musical chords, especially in major and minor consonant triads.

For this purpose, I used math tools of the information theory of emotions that is based on the supposition that positive utilitarian emotions are related with the increase of a certain objective function of the subject, while negative utilitarian emotions are related with the decrease of this function.

The known formulas of emotions were modified by introducing the logarithm of the relative objective function, which, in my opinion, made them closer to reality. A quantitative relation of the force of emotions with generalized parameters of a situation (formula 4) was determined apparently for the first time.

An information theory of emotions of musical chords (outside the context of a musical piece) was created. It defines the sign and the amplitude of the utilitarian emotional coloration of separate major and minor chords through relative pitches of constituent sounds (formula 5).

Limited in scope testing was performed and confirmed, in my opinion, the basic results of the theory. The limits of applicability of the formula of musical emotions within which it correctly renders the sign and, in my opinion, their amplitude were established.

However, complete testing of the theory of chords and the formula of musical emotions requires statistical validation of signs and amplitudes of emotions for the maximum possible amount of major and minor triads and other chords.

## ACKNOWLEDGEMENTS

I express special gratitude to Ernst Terhardt and Yury Savitski for the literature they kindly provided to me for writing this paper.

## APPENDIX

Emotional power *Pwe* of some chords calculated with formula (5).
The most of proportions are direct proportions corresponding to major chords.
*Pwe* of complementary minor chords differ only in sign (see examples).
If *Pwe2* is close in amplitude to *Pwe*, *Pwe2* for side proportions of some chords is provided in brackets. For symmetrical chords of the third group, *Pwe* and *Pwe2* differ only in sign.

```
Main         Side         Pwe (Pwe2)      Commentary
Proportion   Proportion
```

Some symmetrical triads (third group of triads)

```
1:1:1        1:1:1        0     (0)       unison
1:2:4        /4:/2:1      1.    (-1.)
4:6:9        /9:/6:/4     2.58  (-2.58)   "fifth" triad

16:20:25     /25:/20:/16  4.32  (-4.32)   augmented triad
```

Some consonance triads

```
1:2:3        /6:/3:/2     0.86  (-1.72)
2:3:4        /6:/4:/3     1.53  (-2.06)
```

| | | | | |
|---|---|---|---|---|
| 2:3:5 | /15:/10:/6 | 1.64 | | |
| 2:3:8 | /12:/8:/3 | 1.86 | | |
| 2:4:5 | /10:/5:/4 | 1.77 | | |
| 2:5:6 | /15:/6:/5 | 1.97 | | |
| 2:5:8 | /20:/8:/5 | 2.11 | | |
| | | | | |
| 3:4:5 | /20:/15:/12 | 1.97 | | |
| /3:/4:/5 | 20:15:12 | -1.97 | | |
| | | | | |
| 3:4:6 | /4:/3:/2 | -1.53 | (2.06) | |
| 3:4:8 | /8:/6:/3 | 2.19 | (-2.39) | *Pwe2* is close to *-Pwe* |
| 3:5:6 | /10:/6:/5 | 2.16 | (-2.74) | |
| 3:5:8 | /40:/24:/15 | 2.30 | | |
| 3:6:8 | /8:/4:/3 | 2.39 | (-2.19) | *Pwe2* is close to *-Pwe* |
| | | | | |
| 4:5:6 | /15:/12:/10 | 2.30 | | major triad |
| /4:/5:/6 | 15:12:10 | -2.30 | | minor triad |
| | | | | |
| 4:5:8 | /10:/8:/5 | 2.44 | (-2.88) | |
| 5:6:8 | /24:/20:/15 | 2.64 | | |

The author is looking for an organization and financing to continue theoretical and experimental research of this and other important topics.

# Информационная теория эмоций музыкальных аккордов.

## Вадим Р. Мадгазин

*Памяти моих родителей*

*На основании информационной теории эмоций в работе предложено решение многовековой загадки - почему мажорные аккорды воспринимаются как радостные, а минорные - как грустные.*
*Создана соответствующая теория и формула музыкальных эмоций, определяющая знак и амплитуду утилитарной эмоциональной окраски отдельно взятых мажорных и минорных аккордов через относительные высоты составляющих их звуков [15].*

*Ключевые слова: аккорд, мажор, минор, формула музыкальных эмоций, информационная теория эмоций.*

Рисунок 1. Пространство Мажорных и Минорных триад октавы.
Интенсивность точек = const/log(n1*n2*n3) где n1, n2, n3 = 1, 2, 3, ...130

Рисунок 2. Мажорные и Минорные триады и Диады.

# 1. ПРОПОРЦИИ АККОРДОВ

Мы будем изучать психологические эффекты от прослушивания нескольких музыкальных звуков, взятых одновременно, вне любого другого музыкального контекста. Наше внимание будет сконцентрировано на взаимном соотношении высот основных тонов звуков отдельно взятого аккорда. Это соотношение высот мы будем называть пропорцией.

Диада (два звука) в зависимости от пропорции способна производить у субъекта приятное (консонанс) или неприятного (диссонанс) эстетическое впечатление. Считается, что консонантные пропорции состоят целиком из чисел 1,2,3,4,5,6,8.

У триады (аккорда из трёх звуков) появляется новое качество, способность вызывать у субъекта различные эмоции. Эти "музыкальные" эмоции подразделяются на эстетические и утилитарные ("печали" и "радости") [11,12].
Эмоции отдельных аккордов не зависят от изменений общей высоты, громкости или тембра составляющих их звуков. Далее мы будем изучать утилитарные эмоции, порождаемые в основном мажорными и минорными триадами.

Мажорная триада (пропорция 4:5:6, примерно передаётся нотами "до,ми,соль") и минорная триада (пропорция /4:/5:/6 с опущенными единицами в числителях, ноты "до,ми-бемоль,соль") соответствуют положительным эмоциям "радости" и отрицательным эмоциям "печали". Эмоции всех других мажорных и минорных аккордов неизменно имеют такие же знаки.

Математически любая триада состоит из трёх диад и одной общей тройной пропорции. Тройная пропорция может быть записана как "прямая" *A:B:C* или как "обратная" */D:/E:/F* из взаимно простых троек чисел *A,B,C* и *D,E,F.*

Эти пропорции триад можно разделить на три группы:
1. меньшая прямая пропорция         $A*B*C < D*E*F$
2. меньшая обратная пропорция       $A*B*C > D*E*F$
3. обе пропорции одинаковы          $A*B*C = D*E*F.$

Новое (эмоциональное) качество триад, не содержащееся в диадах может быть заключено только в этих общих пропорциях, попадающих в одну из трех вышеописанных групп.

Консонантность триад обычно не превышает консонантности их диад. В музыкальной практике существуют четыре основных типа триад: мажорная и минорная (консонансы), увеличенная и уменьшенная (диссонансы). Триады-консонансы в основном попадают в первые две группы.

Прямые и обратные пропорции увеличенной и уменьшенной триад одинаковы и они попадают в третью группу:
/25:/20:/16=16:20:25
/36:/30:/25=25:30:36.

Все мажорные триады попадают в первую группу, а все минорные - во вторую.
Если некоторая триада *A:B:C* мажорная, то обратная ей триада */C:/B:/A* минорная (хотя все числа триад одинаковы, относительные высоты их средних звуков будут разными). Каждые две такие триады образуют комплементарную пару аккордов, состоящих из одинаковых диад.

Любую прямую пропорцию всегда можно представить в виде обратной, и наоборот:
4:5:6=/15:/12:/10
/4:/5:/6=15:12:10.

С учетом весьма развитого слухового аппарата человека [8,9,14] можно предположить, что хотя представить минорную триаду в виде прямой пропорции (15:12:10) высшей нервной системе субъекта вполне по силам, но также ей по силам представить эту же триаду в виде обратной пропорции (/4:/5:/6). И при этом определить, что данная триада относится ко второй группе, минорной. Такая же классификация может быть произведена как для любой другой консонантной триады, так и для ряда относительно простых диссонантных триад.

Будем называть далее главной пропорцией аккорда ту из двух пропорций высот его звуков (прямую или обратную), которая состоит из меньших чисел (в смысле их произведения), другую же пропорцию будем называть побочной.

Главной пропорцией мажорной триады всегда будет прямая пропорция, а минорной - обратная пропорция. Эмоциональная "информация" аккорда должна определяться однозначно типом его главной пропорции. Другого источника этой информации, обязанной быть в аккорде, просто нет. Нет и не может быть, т.к. мы рассматриваем стационарные аккорды с неизменным спектром, в предельном случае состоящие из чистых тонов.

## 2. ТЕОРИИ МАЖОРА И МИНОРА

Еще в 1558 г. Джозеффо Царлино знал противоположное значение мажорного и минорного аккордов [1]. Однако и 450 лет спустя музыка ушла недалеко от этого знания.

Вот фраза из докторской диссертации по музыкологии 2008 г., по-видимому ставящая жирную точку в вопросе об известных теориях мажора и минора [13, стр. 91]:
"несмотря на то, что многими авторами было описано восприятие мажорных/минорных аккордов и гамм, по-прежнему остается загадкой почему мажорные аккорды ощущаются как радостные, а минорные как грустные".

Но все же не стоит забывать историю. Идея о том, что "смысл" аккорда надо искать вне "собственного пространства" музыки впервые прозвучала как минимум еще сто с лишним лет назад. Цитирую [3,4]:

Гуго Риман к концу своей деятельности отказался от обоснования мажора и консонанса посредством явления обертонов и встал на психологическую точку зрения Штумпфа, рассматривая обертоны лишь как "пример и подтверждение", но не доказательство.
Карл Штумпф перенёс научное обоснование теории музыки из области физиологии в область психологии. Штумпф отказывался объяснять консонанс как акустический феномен, а исходил из психологического факта "слияния тонов".

Но и после этого потребовался целый долгий век для достижения психологией необходимого уровня сложности - появления соответствующего матаппарата. Ведь сколько в любой науке математики, настолько она и точна.

Теперь мы должны подойти к проблеме мажора и минора "с другого конца" и задать вопрос: а что же такое есть эмоция?

# 3. ФОРМУЛЫ ЭМОЦИЙ

На сегодняшний день существует единственная теория эмоций, на языке которой можно надеяться математически описать структуру психологических явлений восприятия музыки. Это информационная теория эмоций [5,6,7,10].

Зависимость эмоции от её объективных параметров называют формулой эмоций. Предположим, что субъект обладает некоторой потребностью величиной $W$. Если ему удаётся получить полезный ресурс $R$, удовлетворяющий эту потребность, то его эмоция $E$ будет положительной, а в случае потери $R$ эмоция будет отрицательной [7]:

$E=F(W,R)$   (1)

Для конкретности можно представить человека, играющего в азартную игру.
Если игрок выиграл некую сумму $R>0$, то у него возникает положительная эмоция радости силой $E$. Если игрок "выиграл" сумму $R<0$ (т.е. проиграл), то возникает отрицательная эмоция горя силой $E$.

Другой подход рассматривает эмоции как средство оптимального управления поведением, направляющее субъекта к достижению максимума его "целевой функции" $P$ [10].
Увеличение функции $P$ сопровождается положительными эмоциями, уменьшение - отрицательными эмоциями.

Поскольку $P$ зависит в простейшем случае от некоторой переменной $x$, то эмоции $E$ вызываются изменением этой переменной от времени [10]:

$E=dP/dt=(dP/dx)*(dx/dt)$   (2)

Отметим, что формулы (1,2) весьма похожи, если учесть что параметр $R$ в (1) на самом деле является разностью между текущим и предыдущим значением некоторого интегрального полезного ресурса $S$, пропорционального целевой функции субъекта.
Например, в случае нашего азартного игрока в качестве $S$ логично выбрать его суммарный капитал, тогда $R=S1-S0=dS=dP$.

Давайте усовершенствуем формулы эмоций, записав их в относительных величинах. В самом деле, радость игрока должна быть пропорциональна относительному размеру выигрыша, а не абсолютному. Учитывая закон Вебера-Фехнера мы можем предположить также, что величина ощущения ($E$) пропорциональна логарифму некоторого "порождающего стимула".

Запишем модифицированную формулу эмоций следующим образом:

$E=F(W)*log(S1/S0)$,   (3)

где $F(W)$ - вынесенная отдельно зависимость эмоций от параметра потребности $W$,
$S1$ - текущее значение целевой функции $S$ (или суммарного полезного ресурса),
$S0$ - значение $S$ в момент времени, предшествующий ситуации, вызвавшей эмоцию.

Разделим обе части (3) на $F(W)$, получив "удельную мощность эмоций на единицу потребности" $Pwe=E/F(W)$. Введём также безразмерную величину $L=S1/S0$, которую логично назвать относительной целевой функцией. Тогда:

*Pwe=E/F(W)=log(S1/S0)=log(L)*,   (4)

Краткое название *Pwe* - "мощность эмоций". *Pwe* пропорциональна затратам эмоциональной энергии в единицу времени - обиходному смыслу выражения "сила эмоций". Измерение силы эмоций в единицах мощности, выделяемой организмом на эмоциональное поведение субъекта предложено в [5].

Легко заметить, что формула (4) даёт верный знак эмоций, положительный при росте *S* (когда *S1>S0* и *L>1*) и отрицательный при падении *S* (когда *S1<S0* и *L<1*).

Применим новую формулу эмоций (4) к восприятию музыкальных аккордов.

## 4. ИНФОРМАЦИОННАЯ ТЕОРИЯ АККОРДОВ

Я предполагаю, что некоторые довольно простые аналогии (на уровне совпадения оценок больше/меньше, лучше/хуже) со смыслом похожей информации из других сенсорных каналов восприятия (визуальных и др.) позволяют психике субъекта классифицировать мажорные аккорды как сигналы, несущие информацию о росте/выгоде/удаче, сопровождаемую положительными эмоциями, а минорные - о падении/убытке/неудаче, сопровождаемую отрицательными.

Т.е. на языке формулы эмоций (4) в мажорном аккорде должна быть заключена информация о событийно-относительной целевой функции *L>1*, а в минорном - о *L<1*.

<u>Моя основная гипотеза состоит в следующем. При восприятии отдельного мажорного или минорного аккорда в психике субъекта порождается значение относительной целевой функции *L*, которое непосредственно связано с главной пропорцией высот его звуков. При этом мажорному аккорду соответствует представление о росте целевой функции (*L>1*), сопровождаемое положительными утилитарными эмоциями, а минорному аккорду соответствует представление о падении целевой функции (*L<1*), сопровождаемое отрицательными утилитарными эмоциями.
 Если главная и побочная пропорции аккорда одинаковы, то он не относится к категории мажора и минора и утилитарная компонента его эмоций отсутствует.</u>

В согласии с этой гипотезой триады из третьей группы отличаются от первых двух (мажора и минора) характером эмоционального отклика субъекта - они вызывают только эстетические эмоции, не имеющие знака.

Запишем формулу (4) для произвольного количества голосов аккорда *M*. Для этого определим *L* как среднее геометрическое всех чисел главной пропорции аккорда, получив в итоге <u>окончательный вид "формулы музыкальных эмоций"</u>:

*Pwe=log(L)=(1/M)\*log(n1\*n2\*n3\* ... \*nM)*,   (5)

где *ni* - это *i*-й голос аккорда в виде соответствующего целого числа или обратной дроби из главной пропорции высот голосов аккорда, а логарифм берётся по основанию 2.

Из (5) непосредственно следует, что *Pwe* мажорной триады будет совпадать по амплитуде с

*Pwe* комплементарной ей минорной триады, отличаясь только знаком.

Логичное название для *Pwe* (5) - "эмоциональная мощность аккорда", или просто "мощность аккорда", положительная для мажора и отрицательная для минора (аналогия: приток и отток эмоциональной энергии).

## 5. ОБСУЖДЕНИЕ РЕЗУЛЬТАТОВ

Итак, выдвинув ряд довольно простых предположений, мы получили новые формулы (3,4,5), которые связывают обобщенные параметры ситуации (или конкретные параметры аккордов) со знаком и силой вызываемых ими в контексте ситуации утилитарных эмоций.

Как можно оценить этот результат?
Цитирую [5]:
"Попыток объективного определения силы эмоции, вероятно, не было. Однако можно предположить, что такое определение должно быть основано на энергетических представлениях. Если эмоция вызывает некоторое поведение, то это поведение требует определенного расхода энергии. Чем сильнее эмоция, тем интенсивнее поведение, тем больше требуется энергии в единицу времени.
Т.е. силу эмоции можно попытаться отождествить с величиной мощности, которую организм выделяет на соответствующее поведение."

Попробуем максимально критически подойти к нашим новым результатам, раз их пока что не с чем сравнить.

Во первых, мощность эмоций *Pwe* из формул (4,5) хотя и пропорциональна "субъективной силе" эмоций, но их связь может быть не линейной. И эта связь - лишь некая средняя зависимость по всему континууму субъектов и может быть подвержена индивидуальным отклонениям. Например, возможно, что вместо среднего геометрического в формуле (5) следует применить какую-то другую функцию от чисел пропорции аккорда.

Во-вторых, если иметь ввиду конкретный вид формулы музыкальных эмоций (5), то следует заметить, что хотя формально в ней *M* и может быть равно 1 или 2, мы имеем право говорить о возникновении утилитарных эмоций только при *M>2*. Однако уже при *M=2* у нас возникают эстетические эмоции, а при *M>3*, возможно появления дополнительных факторов, как-то влияющих на результат.
Весьма показательным является тот факт, что для случая M=2 формула (5) с точностью до коэффициента 1/2 совпадает с HD-функцией (harmonic distance), предложенной в [16] для анализа музыкальных интервалов.

В третьих, по-видимому область валидных значений амплитуды *Pwe* для категории мажора и минора имеет верхнюю границу 2.7...3.0, но где-то уже со значения 2.4 начинается область насыщения утилитарно-эмоционального восприятие аккордов, и примерно там же проходит нижняя граница диапазона возможного "вторжения" диссонансов.
Но это последнее - общая проблема "не монотонности" ряда диссонантных интервалов, не связанная напрямую с эмоциональным аспектом восприятия аккордов. А ограниченность динамического диапазона мощности эмоций - общее свойство любой сенсорной системы человека, легко объясняемое отсутствием аналогий в "реальной жизни", соответствующих слишком сильным изменениям целевой функции - в 10 раз и более за одно событие.

**15**

В четвёртых, аккорды, у которых прямые и обратные пропорции состоят из одинаковых (или почти одинаковых) чисел - не имеют (или почти не имеют) утилитарной компоненты эмоций, что соответствует нулевой (или почти нулевой) *Pwe*. Однако в данном случае формула (5) выдаст другой, неверный результат.

Тут проявляется весьма простое явление: раз невозможно отличить по сложности прямую и обратную пропорции аккорда, то субъект не может определить направление изменения целевой функции и классификация этого аккорда в категориях утилитарных эмоций просто не производится.

Можно дополнить формальный результат применения формулы (5) дополнительным правилом. Для каждого аккорда определяется два значения: *Pwe* по главной пропорции аккорда и *Pwe2* по побочной пропорции аккорда.

Если *Pwe* и *Pwe2* близки по амплитуде, то утилитарные эмоции будут пропорциональны их полусумме (т.е. почти нулевые):

*E=(Pwe+Pwe2)/2*,   (5а)

И это правило по-видимому начинает работать при модуле *(Pwe+Pwe2)*, меньшем чем 0.50.

Однако при всех своих мнимых или действительных недостатках формулы (3,4,5 и 5а) все же дают нам полезный материал для численных оценок силы эмоций, которые довольно просто сделать по крайней мере в одной конкретной области - области эмоционального восприятия отдельных мажорных и минорных аккордов.

Прослушаем мажорные триады 2:3:5 и 3:4:5. На слух мне кажется вполне заметным, что *Pwe* первой триады (1.64) действительно несколько меньше *Pwe* второй (1.97).
Примерно то же самое происходит и с соответствующими парами минорных триад.

То же самое можно проделать с любой другой парой консонантных триад (из 1-й или 2-й, но не из 3-й группы), на мой взгляд - с аналогичным результатом: мощность (сила, яркость, глубина) эмоций кажется больше у той триады, у которой больше амплитуда *Pwe*.

Это впечатление пропорциональности силы эмоций и *Pwe* сохраняется и при прослушивании аккордов, составленных из звуков с богатым гармоническим спектром.

## 6. ВЫВОДЫ

Целью моей работы являлось установление возможной природы эмоциональной окраски музыкальных аккордов, в особенности консонантных мажорных и минорных триад.

Для этого я использовал матаппарат информационной теории эмоций, базирующейся на предположении о том, что положительные утилитарные эмоции связаны с ростом некоторой целевой функции субъекта, а отрицательные - с её падением.

Известные формулы эмоций были модифицированы введением логарифма относительной целевой функции, что по моему мнению сделало их более соответствующими реальности. По-видимому впервые определена количественная связь силы эмоций с обобщенными параметрами ситуации (формула 4).

Создана информационная теория эмоций музыкальных аккордов (вне контекста музыкального произведения) и предложена формула музыкальных эмоций (5), определяющая знак и амплитуду утилитарной эмоциональной окраски отдельно взятых консонантных мажорных и минорных аккордов через относительные высоты составляющих их звуков.

Выполнена ограниченная по объему проверка, по моему мнению подтверждающая основные результаты теории. Описаны пределы применимости формулы музыкальных эмоций, в которых она верно передает знак и (на мой взгляд) их амплитуду.

Однако для более полной проверки теории аккордов и формулы музыкальных эмоций требуется статистически достоверное определение знаков и амплитуд эмоций для максимально возможного количества мажорных и минорных триад и других аккордов.

## 7. ССЫЛКИ

## БЛАГОДАРНОСТИ

## ПРИЛОЖЕНИЕ

Эмоциональная мощность *Pwe* некоторых аккордов, рассчитанная по формуле (5).
Основная часть пропорций - прямые пропорции, соответствующие мажорным аккордам.
*Pwe* комплементарных им минорных аккордов отличаются только знаком (см. примеры).
В скобках дана *Pwe2* для побочных пропорций некоторых аккордов, если *Pwe2* по амплитуде приближается к *Pwe*. Для симметричных аккордов третьей группы *Pwe* и *Pwe2* отличаются только знаком.

| Главная пропорция | Побочная пропорция | *Pwe* (*Pwe2*) | | Примечание |
|---|---|---|---|---|

Некоторые симметричные триады (третья группа триад)

| | | | | |
|---|---|---|---|---|
| 1:1:1 | 1:1:1 | 0 | (0) | унисон |
| 1:2:4 | /4:/2:1 | 1. | (-1.) | |
| 4:6:9 | /9:/6:/4 | 2.58 | (-2.58) | "квинтовая" триада |
| 16:20:25 | /25:/20:/16 | 4.32 | (-4.32) | увеличенная триада |

Некоторые консонантные триады

| | | | | |
|---|---|---|---|---|
| 1:2:3 | /6:/3:/2 | 0.86 | (-1.72) | |
| 2:3:4 | /6:/4:/3 | 1.53 | (-2.06) | |
| 2:3:5 | /15:/10:/6 | 1.64 | | |
| 2:3:8 | /12:/8:/3 | 1.86 | | |
| 2:4:5 | /10:/5:/4 | 1.77 | | |
| 2:5:6 | /15:/6:/5 | 1.97 | | |
| 2:5:8 | /20:/8:/5 | 2.11 | | |
| | | | | |
| 3:4:5 | /20:/15:/12 | 1.97 | | |
| /3:/4:/5 | 20:15:12 | -1.97 | | |
| | | | | |
| 3:4:6 | /4:/3:/2 | -1.53 | (2.06) | |
| 3:4:8 | /8:/6:/3 | 2.19 | (-2.39) | *Pwe2* близко к -*Pwe* |
| 3:5:6 | /10:/6:/5 | 2.16 | (-2.74) | |
| 3:5:8 | /40:/24:/15 | 2.30 | | |
| 3:6:8 | /8:/4:/3 | 2.39 | (-2.19) | *Pwe2* близко к -*Pwe* |
| | | | | |
| 4:5:6 | /15:/12:/10 | 2.30 | | мажорная триада |
| /4:/5:/6 | 15:12:10 | -2.30 | | минорная триада |
| | | | | |
| 4:5:8 | /10:/8:/5 | 2.44 | (-2.88) | |
| 5:6:8 | /24:/20:/15 | 2.64 | | |

Copyright (c) Vadim R. Madgazin, 2009

Версия текста от 10 сентября 2011 г.